\documentclass[runningheads]{llncs}
\usepackage{graphicx}
\usepackage{multirow}
\usepackage{float}
\usepackage{caption}
\usepackage{subcaption}
\usepackage[ruled, linesnumbered]{algorithm2e}
\usepackage{multicol}
\usepackage{lipsum}  
\usepackage{hyperref}
\usepackage{soul}
\usepackage{booktabs}
\usepackage{perpage} %the perpage package
\MakePerPage{footnote}

\SetKwComment{Comment}{/* }{*/}
\newcommand{\etal}{\textit{et al. }}

\usepackage{color}
\newcommand{\ra}[1]{\renewcommand{\arraystretch}{#1}}

\begin{document}
\title{POTHER: Patch-Voted Deep Learning-based Chest X-ray Bias Analysis for COVID-19 Detection}
% \title{POTHER: Patch-Voted COVID-19 Detection}
%
\titlerunning{Chest X-ray Bias Analysis for COVID-19 Detection}
\author{Tomasz Szczepański\inst{1}\orcidID{0000-0001-6189-478X} \and
Arkadiusz Sitek\inst{2}\orcidID{0000-0002-0677-4002} \and
Tomasz Trzciński\inst{1, 3, 4}\orcidID{0000-0002-1486-8906} \and
Szymon Płotka \inst{1, 2}\orcidID{0000-0001-9411-820X}}
% %
\authorrunning{T. Szczepański et al.}
\institute{Warsaw University of Technology, Warsaw, Poland \and
Sano Centre for Computational Medicine, Cracow, Poland \and
Jagiellonian University, Cracow, Poland \and
Tooploox, Wroclaw, Poland}
\maketitle              % typeset the header of the contribution
\begin{abstract}
A critical step in the fight against COVID-19, which continues to have a catastrophic impact on peoples lives, is the effective screening of patients presented in the clinics with severe COVID-19 symptoms. Chest radiography is one of the promising screening approaches. Many studies reported detecting COVID-19 in chest X-rays accurately using deep learning. 
A serious limitation of many published approaches is insufficient attention paid to explaining decisions made by deep learning models. 
Using explainable artificial intelligence methods, we demonstrate that model decisions may rely on confounding factors rather than medical pathology. After an analysis of potential confounding factors found on chest X-ray images, we propose a novel method to minimise their negative impact. We show that our proposed method is more robust than previous attempts to counter confounding factors such as  ECG leads in chest X-rays that often influence model classification decisions. In addition to being robust, our method achieves results comparable to the state-of-the-art. The source code and pre-trained weights are publicly available at (\url{https://github.com/tomek1911/POTHER}). 

\end{abstract}
\section{Introduction}

The SARS-CoV-2 outbreak has claimed the lives of millions of people, and despite its onset in 2019, it remains a serious concern and a threat to public health. The gold standard for diagnosing COVID-19 disease is an RT-PCR test; however, it is expensive, necessitates specialised laboratories and requires the patient to wait relatively long for the outcome. For this reason, computer scientists and radiologists become interested in the computer-aided diagnosis (CAD) capabilities of chest X-rays (CXR). The automatic diagnosis of COVID-19 using chest X-ray images is challenging due to the high intra-class variations, superimposition of anatomical structures, or implanted electronic devices \cite{Calli}. Moreover, a significant limitation in developing reliable models for detecting pneumonia and COVID-19 is the lack of precisely annotated and rigorously collected data. These limitations may result in models learning various confounding factors in the CXRs. We investigate this phenomenon in our work.\\
\indent
Most of the approaches developed for the classification of the CXR images relay on global features \cite{Minaee}, \cite{alwaisy}, \cite{rajaraman}, \cite{ucar}. However, these features may not accurately represent the complex nature of CXR images \cite{cabrera}. It should be verified when working with medical images, especially when the model's accuracy is very high \cite{pham}, \cite{nayak}, the reasons why algorithms perform so well to prevent developing algorithms which base their decision on confounding factors rather then medical pathology.
Wang \etal \cite{COVID_Net} propose COVID-Net, the first lightweight capacity neural network dedicated for COVID-19 detection and introduce a novel COVIDx dataset. Authors create a feature extraction method tailored for COVID-19 and provide results of the model's decisions analysis with the use of GSInquire \cite{lin}. They create a map of lung areas important for COVID-19 detection and claim a production-ready solution. Considering the need for a critical approach to chest classification and analysis of biases\cite{maguolo},\cite{cabrera},\cite{degrave}, we carry out further research into the analysis of the deep learning model decisions using the COVIDx dataset.\\
\indent
Recently, a new patch-based learning technique \cite{Li}, \cite{oh}, \cite{roy} emerged as a successful method for robust model learning and generalization. Li \etal \cite{Li} propose a multi-resolution patch-based CNN approach for lung nodule detection on CXR image. The method achieves high accuracy by leveraging the local context of CXR images. Roy \etal \cite{roy} use the model's local training method for classification on the ICIAR-2018 breast cancer histology image dataset, which allows them to achieve state-of-the-art results on this dataset. Oh \etal \cite{oh}, inspired by their statistical analysis of the potential imaging biomarkers of the CXRs, explore patch-based learning for COVID-19 detection. They propose random patch cropping, from which the final classification result is obtained by majority voting from inference results at multiple patch locations. Authors analyse model decisions using proposed probability gradient-weighted class activation map (GradCAM) \cite{GradCAM} as a part of explainable AI (XAI) method \cite{xai}, upon which they conclude their results are correlated with radiological findings. Similar to Oh \textit{et al.}, our proposed method POTHER explores a patch-based learning approach in CXR images for reliably detecting COVID-19. In contrast, we do not use segmented lungs as input; instead, we propose a multi-task model that leverages segmentation task to extract valuable features. In addition, we limit the area from which we draw patches and reduce their size.\\
\indent
We use XAI methods to demonstrate that model decisions may rely on \textit{confounding factors} rather than medical pathology. Degrave \etal \cite{degrave} and Cabrera \etal \cite{cabrera} call them shortcuts, while Maguolo \etal \cite{maguolo} biases. An  analysis of shortcuts in CXRs, which we call {\em confounding biases} (CBs), based on open-source datasets and global learning methods, is presented by \cite{degrave}. It demonstrates the detrimental influence on models decisions caused by laterality tokens, i.e. L or R letter, meaning the left or right side of the image, the position of the clavicles and the presence of arms in the upper parts of the image. Authors present that machine learning (ML) models trained on CXR images may generalise poorly and owe the majority of their performance to the learning of shortcuts that may be consistently detected in both internal and external domains - making external validation alone insufficient to detect poorly behaved models. We analyse CBs in the frequently cited COVIDx dataset \cite{COVID_Net} and propose a multi-task learning COVID-19 detection method called POTHER that improves robustness to CBs. 
\begin{figure*}[t!]
    \centering
    \includegraphics[width=1.0\textwidth]{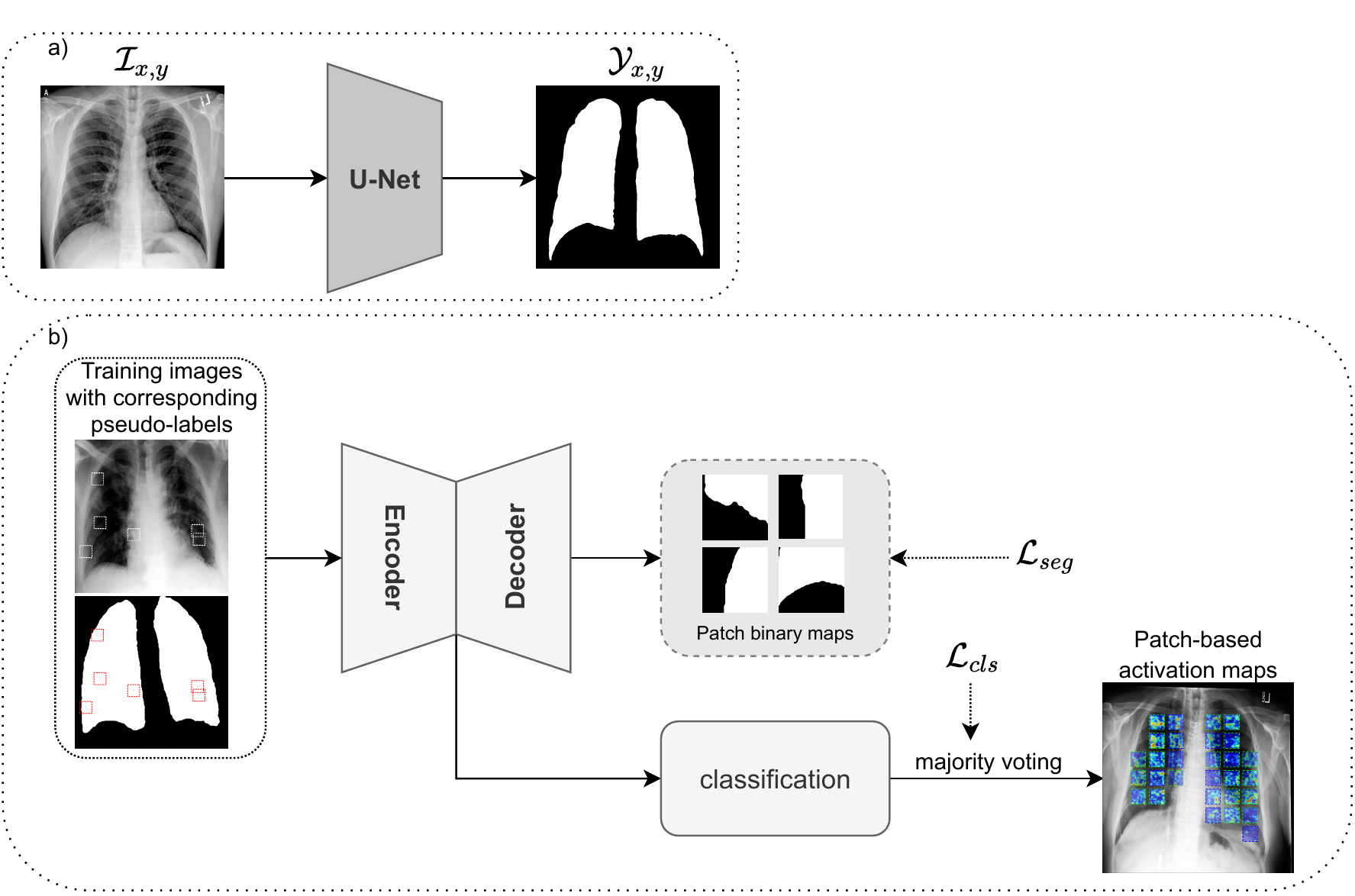}
    \caption{The overview of the proposed POTHER framework for chest X-ray bias analysis. (a) Pre-trained segmentation network to generate pseudo-labels, (b) an encoder-decoder patch-based multi-task learning network to classify CXR images. We adopt majority patch-based voting for classification and employ patch-based activation maps to explain the results.}
    \label{fig:pother}
\end{figure*}\\

The contributions of this paper are threefold. Firstly, we propose a novel multi-task patch-voted learning-based method called POTHER for chest X-ray image analysis for COVID-19 detection. Secondly, we analyse activation maps of currently available methods and sources of confounding biases in CXRs from the COVIDx dataset. Our analysis reveals a significant number of CBs. These biases should be considered when using the COVIDx dataset in future research. The main CBs are ECG leads, laterality tokens and hospital markings. Finally, we demonstrate that deep learning models learn how to classify pneumonia manifestations and lung morphology, focusing on the confounding mentioned above factors rather than the actual manifestation of the disease or lack thereof on lung parenchyma. In our methods, to counter CBs, we used segmentation as a helper task which allowed for an efficient feature extraction method that performs comparably to the state of the art. The paper is organized as follows. Section \ref{method} details our proposed method. Next, Section \ref{experiments} presents details about our experimental setup and results. We follow with Section \ref{discussion} which discusses the results. Finally, Section \ref{conclusions} concludes the paper.

\section{Method} \label{method}

Fig. \ref{fig:pother} presents an overview of our proposed method. First, we pre-train the U-Net lung fields segmentation network to generate pseudo-labels. Second, we use patches to train a multi-task U-Net-based neural network for CXR image bias analysis. Finally, for classification, we use a majority patch-based voting.\\
\noindent
\textbf{Pre-training segmentation network.} We train a segmentation neural network to extract lung fields from CXR images. For this task, we adopt the U-Net encoder-decoder to perform semantic segmentation \cite{unet}. With the pre-trained model on \cite{cxrdataset}, \cite{C19dataset} datasets, we generate pseudo labels for unlabelled COVIDx dataset \cite{COVID_Net}. Before feeding those pseudo masks to the multi-task neural network, we use custom pre-processing algorithms.\\
\noindent
\textbf{Multi-task neural network.} Inspired by \cite{plotka}, \cite{plotka2}, \cite{amyar}, we use an encoder-decoder based convolutional neural network (CNN) for simultaneous classification and segmentation of the lungs in the CXRs. Our network is U-Net-based with ImageNet pre-trained ResNet-50 as backbone encoder \cite{resnet}. We use an encoder part to extract high-level X-ray image features. Every encoder block forwards feature maps and concatenates them with the corresponding decoder part. We employ an attention mechanism to the skip connections to learn salient maps suppressing irrelevant lung vicinity that may be a source of CBs. We extend an attention gate mechanism by aggregating multi-scale feature maps from the decoder to learn the local context of the lung feature maps representation. Each feature map is fed through an inception module that leverages convolutional filters of multiple kernels (i.e. 1x1, 3x3, 5x5, and 7x7) and stride sizes ($S=1, S=2$), which does not increase the number of parameters significantly. The amplitudes of features from the deeper layers are smaller than the shallow ones. To prevent shallow layers from dominating deeper ones, we normalise the weights of the features from multiple scales with the L2 norm before concatenation.\\
\noindent
\textbf{Patch-based learning.} Motivated by \cite{oh}, we adopt a patch-based learning method to train our multi-task neural network. Unlike Oh, we do not cut out the lungs with masks to avoid inductive bias. We use the lung masks instead as pseudo-labels for segmentation training. This helper task requires the model to learn new features necessary to recognise lung tissue boundaries. A pre-processed CXR image and corresponding mask are resized to $1024 \times 1024$. Then using a draw area based on the scaled-down, with a ratio of 0.9, whole lung mask, we randomly choose a patch centre. Its coordinates are drawn with a uniform distribution of non-zero pixels from the draw area mask. Fig. \ref{fig:pother}(b) shows training images with corresponding pseudo-labels in the context of the entire CXR image, i.e. exemplary drawn image patches are marked with white squares in the CXR image and pseudo-labels with red squares on the corresponding mask. In our method, we use patches of size $80 \times 80$, in contrast to Oh uses patches of size $224 \times 224$.
A region of the image and the corresponding mask are cut off using determined patch coordinates, resulting in an image patch and pseudo-label pair. Then the image patch and its mask are interpolated to $224 \times 224$. This pair is used to train a segmentation head, while the image patch and its corresponding class represent a training pair for the classification head. Thanks to the reduction of the draw area and the small size of the patch, the model's input never contains laterality tokens and hospital markings. At the same time, if chosen close to the lung edge, a patch covers a small lung boundary fragment, allowing the model to recognise lung tissue based on the segmentation task.\\
\noindent
\textbf{Majority patch-based voting.} Only one patch per image, per batch, is used in the training phase. However, we repeat the random draw multiple times for the inference to cover the whole lung field as the area of interest. Each time, based on the patch, the model makes a single classification called a single vote. The majority vote result, i.e. the class chosen based on the majority of the patches, is the final classification for that image. 

\section{Experiments and Results} \label{experiments}

In this section, we evaluate our methods on the COVIDx dataset. Next, we present gradient-weighted class activation maps that compare the decision bases of our proposed method with existing methods. We employ GradCAM to provide visual explanations of our method decision bases that focuses its attention on lung morphology and is less sensitive to CBs. Finally, we show quantitative results on the COVIDx test set.

\subsection{Datasets}
To pre-train a neural network for the segmentation of lungs, we use data from \cite{cxrdataset}, \cite{C19dataset}. The datasets consist of 6380 2D CXR images. The annotations for each image provide a manual segmentation mask rendered as polygons, including the retrocardiac region.
To evaluate our methods, we use an open-source \textbf{COVIDx dataset} \cite{COVID_Net}. In total, the dataset consists of 13970 2D CXR images (8806 normal, 5551 pneumonia and 353 COVID-19 cases). The authors constantly expand this dataset, and in order to compare with their published results, we use the same version of the dataset as they used at the time of publication. We create a validation set by splitting the training set with a 70:30 ratio, and the test dataset is built using the script provided by the COVIDx authors.

\subsection{Data pre-processing}
\textbf{Lung masks.} We remove unnecessary objects from lungs masks like electronic devices, which are labelled with a shade of grey. We perform mask filtering to improve the masks or remove them from the dataset when the correction is not possible. We present a detailed algorithm in the supplement \footnote{\url{https://cutt.ly/rIB1JFQ}}. Finally, we resize the original image and the mask to a size of $1024 \times 1024$ pixels using linear interpolation. All images are resized to square regardless of their original aspect ratio.

\noindent
\textbf{COVIDx dataset.} The images for the training of our model are pre-processed with histogram equalisation. 
Soft augmentation methods consist of offset, scaling without preserved aspect ratio, rotation, horizontal flip, Contrast Limited Adaptive Histogram Equalisation (CLAHE) with random clip range and random grid range, brightness and contrast adjustment, sharpening and embossing.  

\subsection{Implementation details}

We implement our model in PyTorch deep learning framework and train on a workstation with a single GPU NVIDIA Titan RTX 24GB until convergence over 100 epochs, with a mini-batch size of 16, an initial learning rate of $ 1 \times 10^{-4}$ and a weight decay factor of $1 \times 10^{-4}$. We set Rectified Adam (RAdam) as the optimiser to minimise the loss function. To prevent overfitting, we apply various soft data augmentation techniques. During training, we perform the following transformations: horizontal flip, sharpen, emboss and CLAHE with $p = 0.5$. We also apply a random-weighted sampler. The weights are computed as an inverse class frequency. As loss function, we use Dice loss for segmentation task:

\begin{equation} 
    \mathcal{L}_{dice} \ =\ 1 - \frac{2 \sum_{i}^{N} p_{i}g_{i} + \epsilon}{\sum_{i}^{N}p_{i}^2 + \sum_{i}^{N}g_{i}^2 + \epsilon},
\end{equation}
\noindent
where $p_{i}$ is the prediction pixel value, $g_{i}$ is the ground truth, and $\epsilon$ is a numerical stability to avoid divide by zero errors, and Weighted Cross-Entropy (WCE) for classification task:

\begin{equation}
    \mathcal{L}_{WCE} \ =\ - \frac{1}{N} \sum_{n=1}^{N} wr_{n} \log(p_{n}) + (1 - r_{n})\log(1 - p_{n}),
\end{equation}
\noindent 
where $w$ is the class weight, $p_{n}$ is the Softmax probability for the $i^{th}$ class, and $r_{n}$ is the ground truth value of $\{0, 1\}$.
Finally, the overall multi-task loss function can be formulated as the sum of both loss functions:

\begin{equation}
    \mathcal{L} = \mathcal{L}_{Dice} + \mathcal{L}_{WCE}.
\end{equation}

% \begin{equation}
%     \mathcal{L} = \lambda \mathcal{L}_{Dice} + \mathcal{L}_{WCE}
% \end{equation}
% where $\lambda$ is the weighting parameter. \tom{The weight is set experimentally to 10, and its purpose is to equalise the convergence rate of both tasks. It makes the model more attentive to segmentation and focuses its attention on the lung tissue.} \arek{What value was used, specify it here. Why? Was there any experimentation done to investigate influence of lambda on results? If not, just leave it}. 

\subsection{Gradient-weighted class activation mapping results}

\begin{figure*}[t!]
   \centering
    \begin{tabular}{ccc}
        \includegraphics[width=3.4cm]{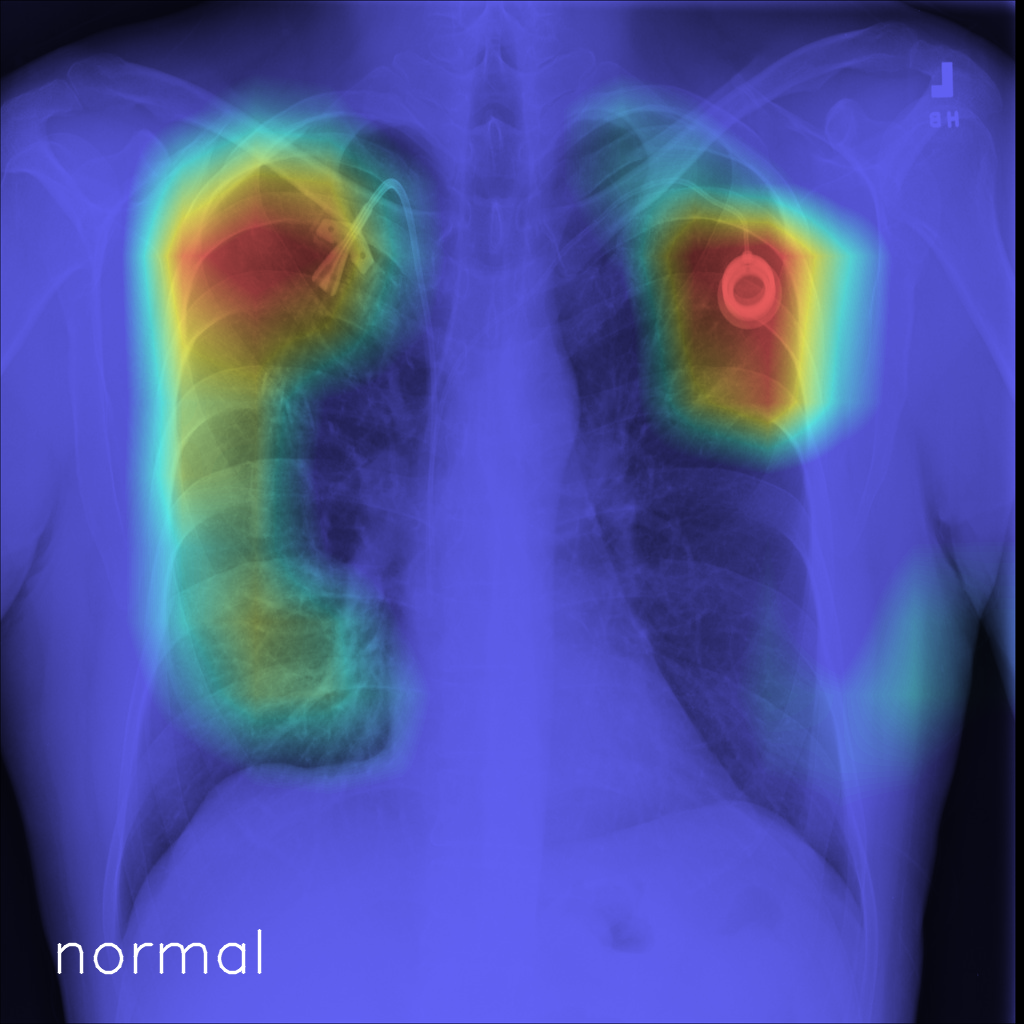}&
        \includegraphics[width=3.4cm]{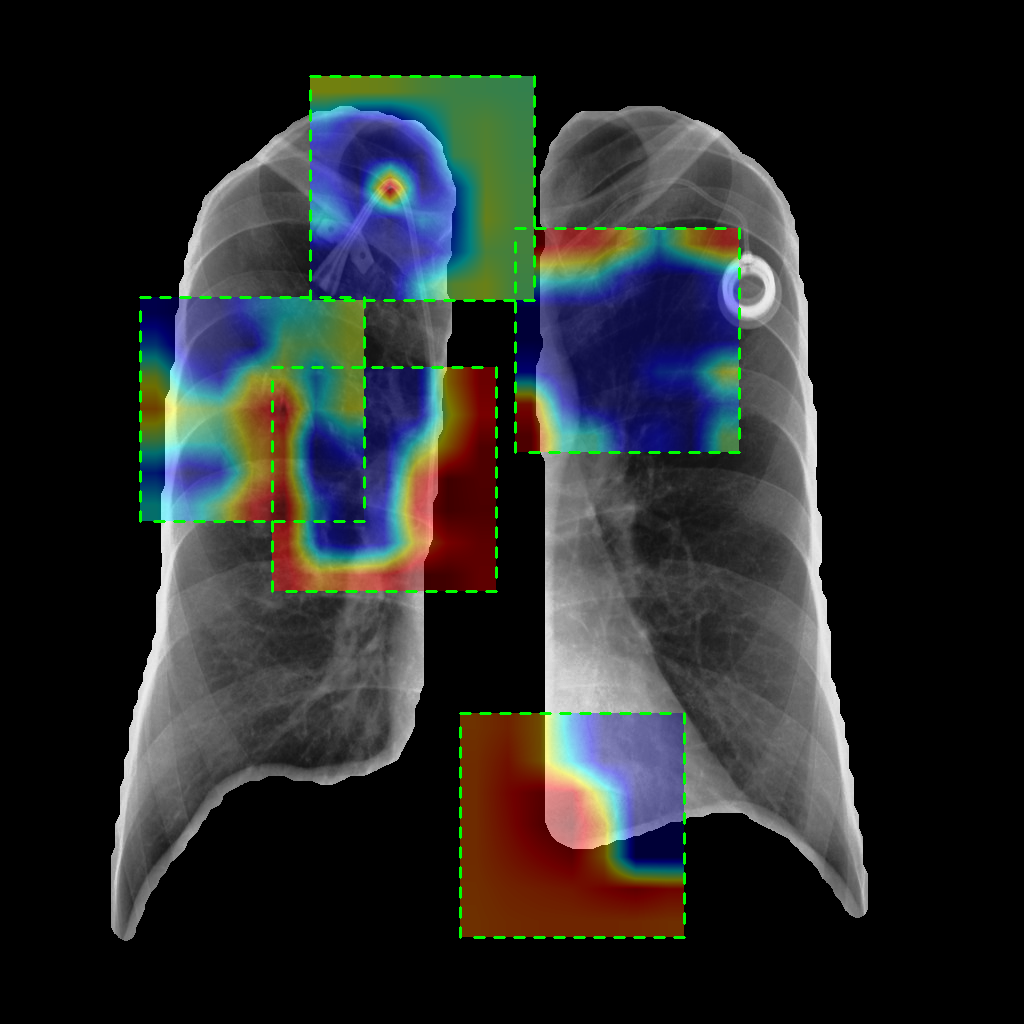}&
        \includegraphics[width=3.4cm]{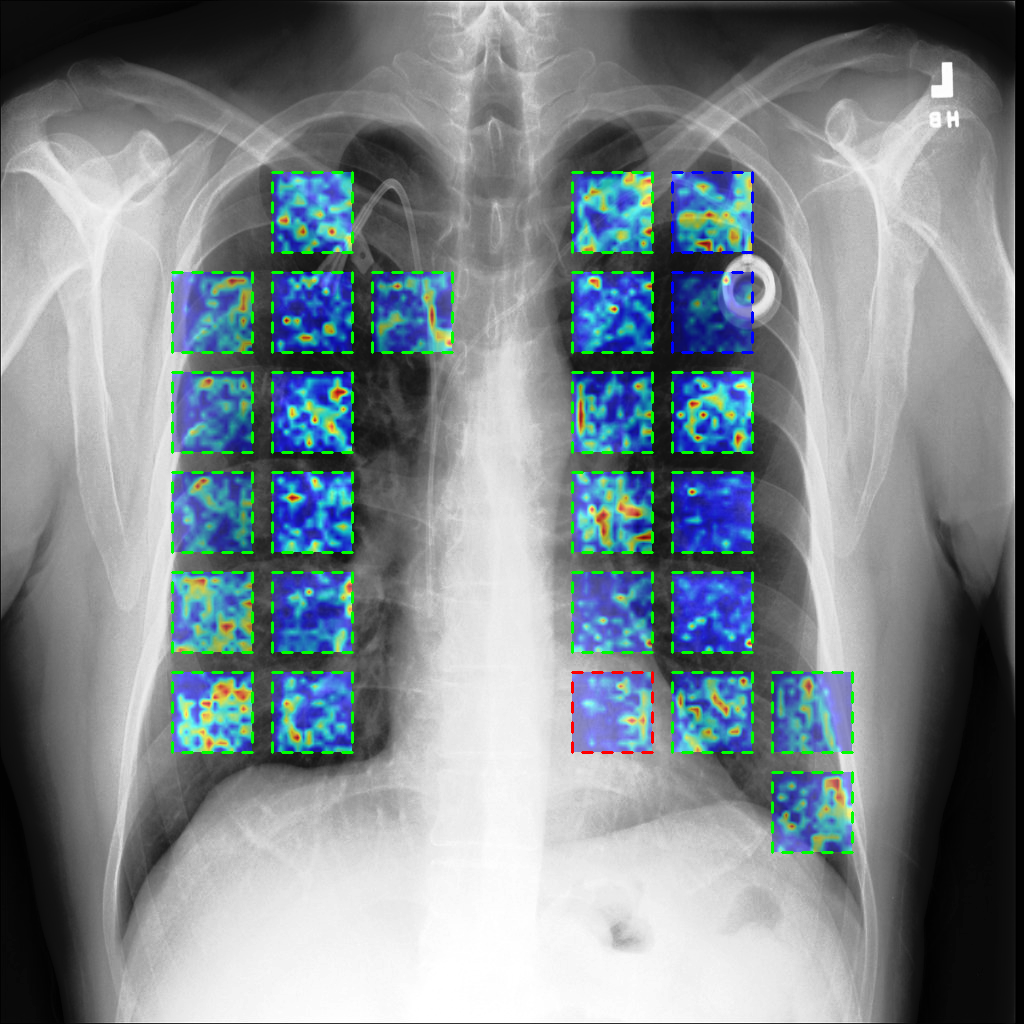} \\
        \includegraphics[width=3.4cm]{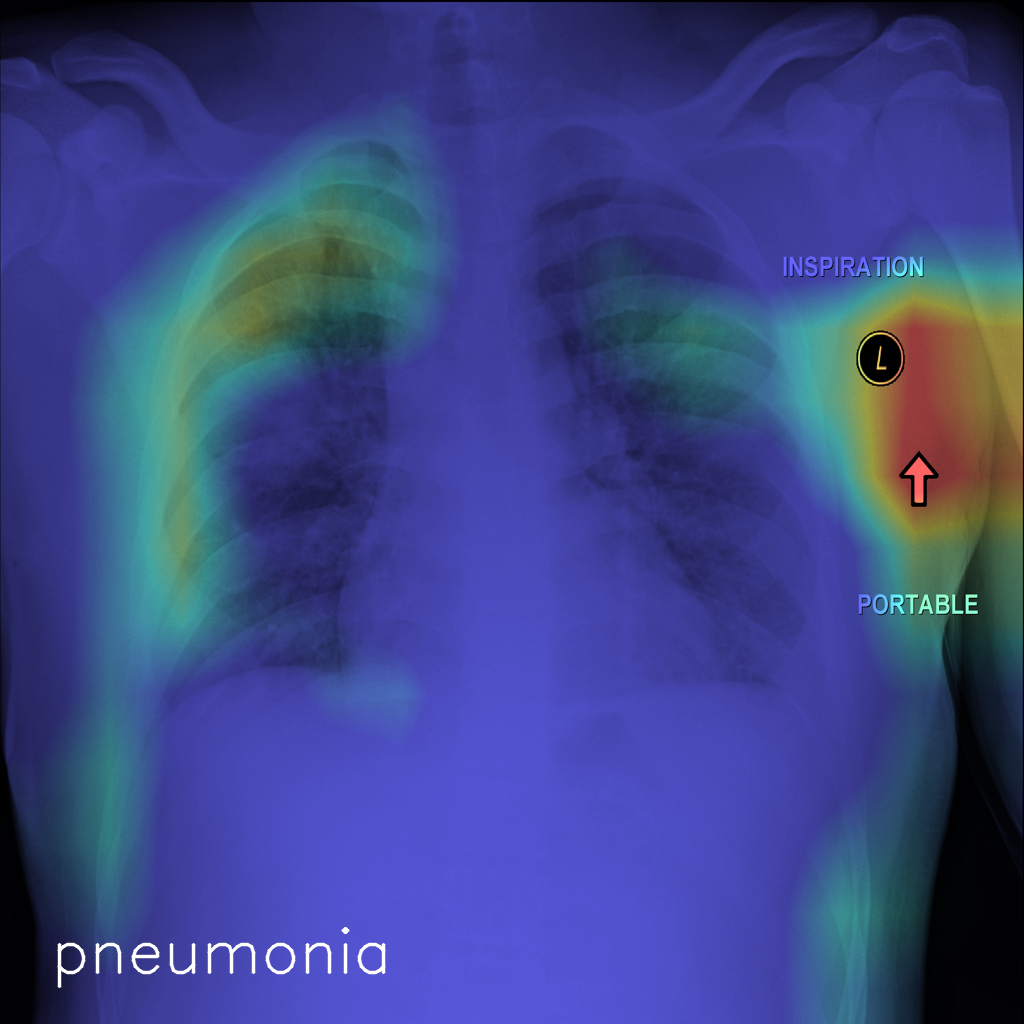}&
        \includegraphics[width=3.4cm]{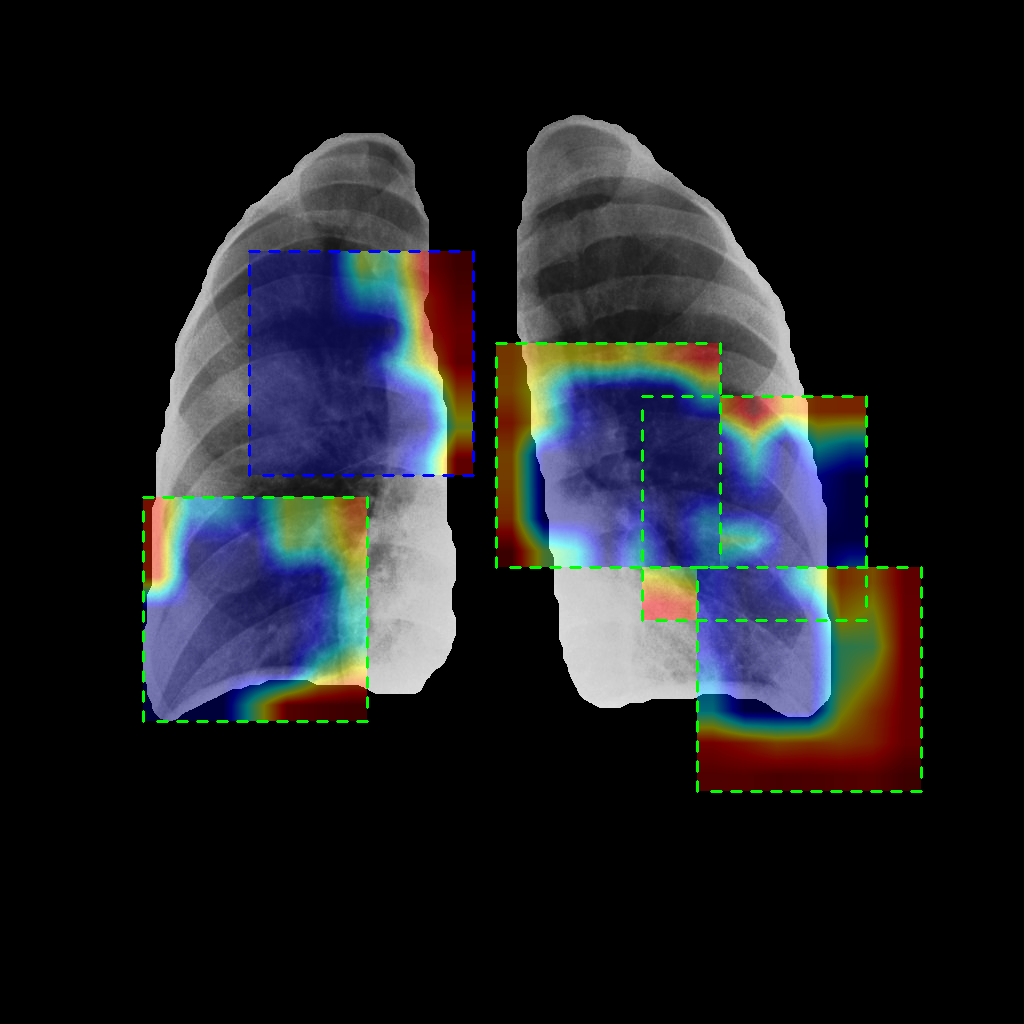}&
        \includegraphics[width=3.4cm]{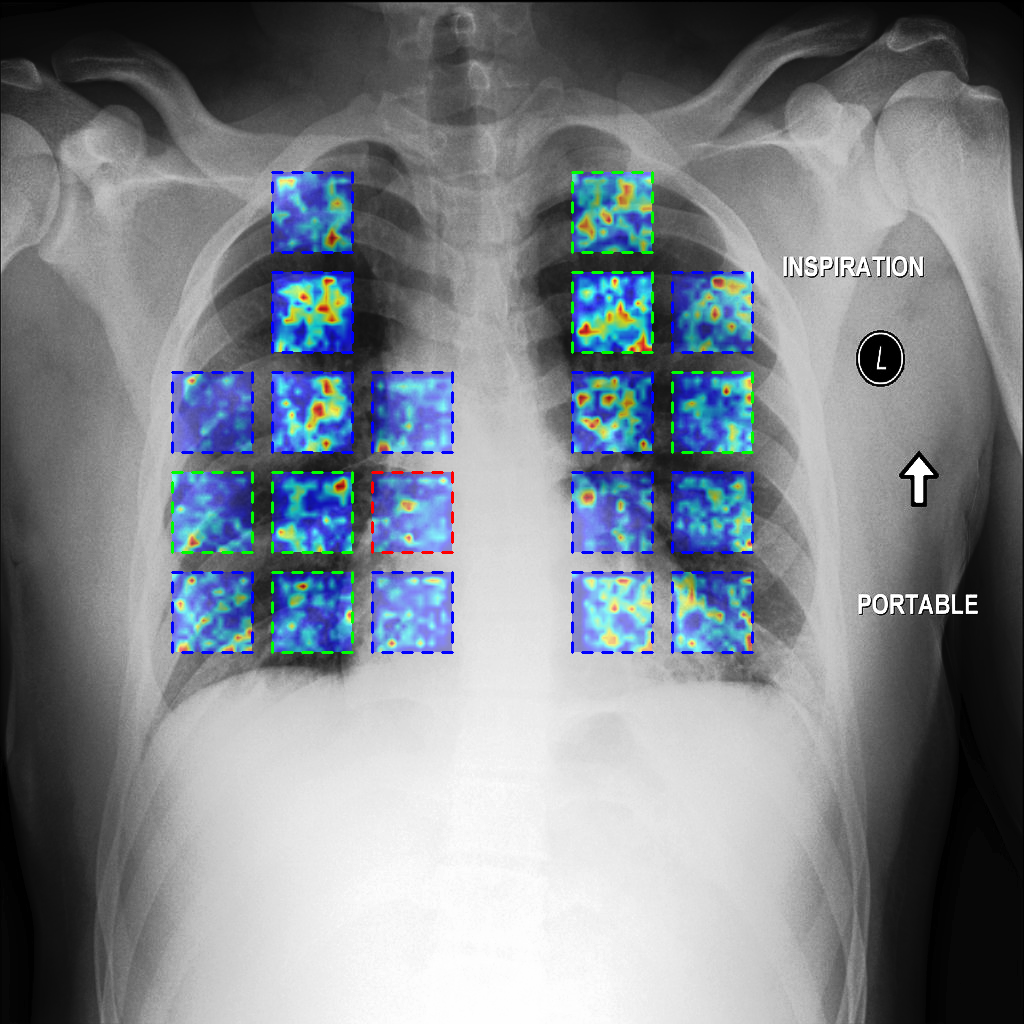} \\
        \includegraphics[width=3.4cm]{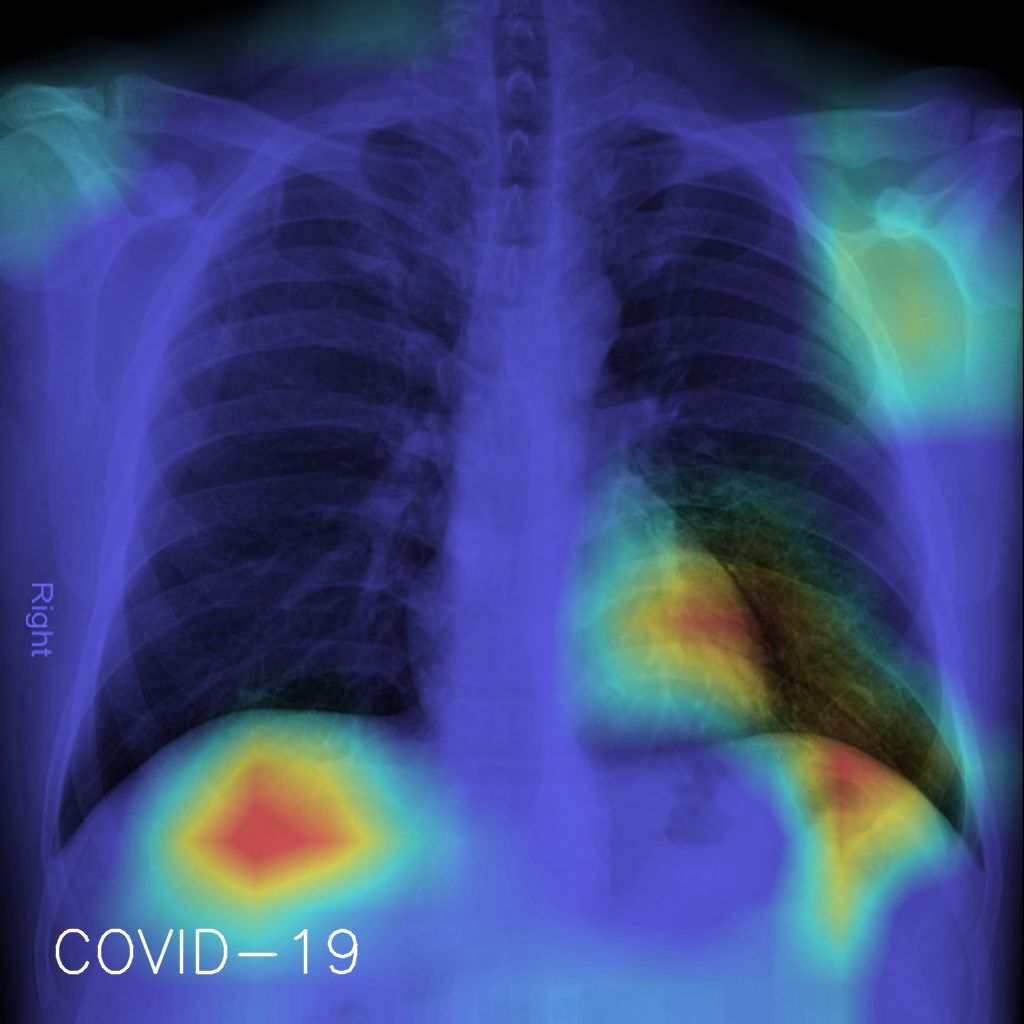}&
        \includegraphics[width=3.4cm]{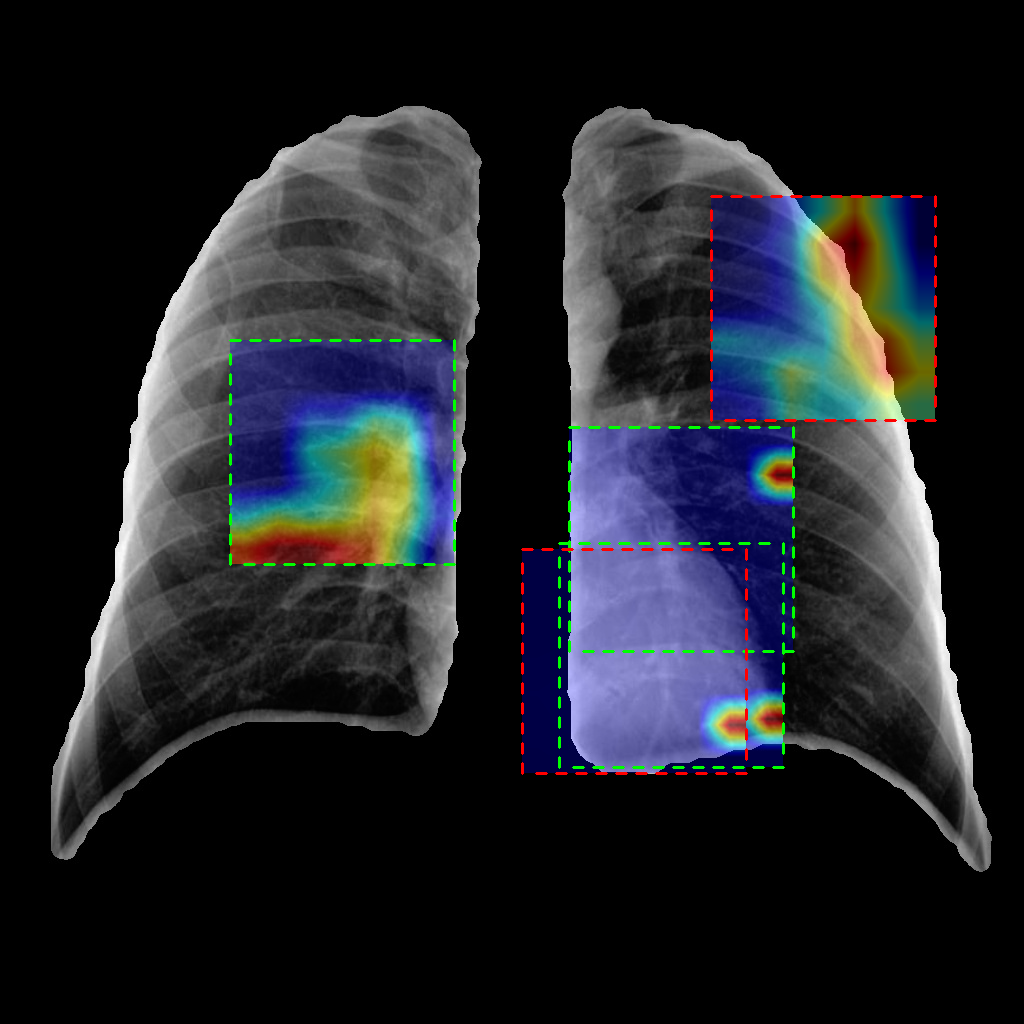}&
        \includegraphics[width=3.4cm]{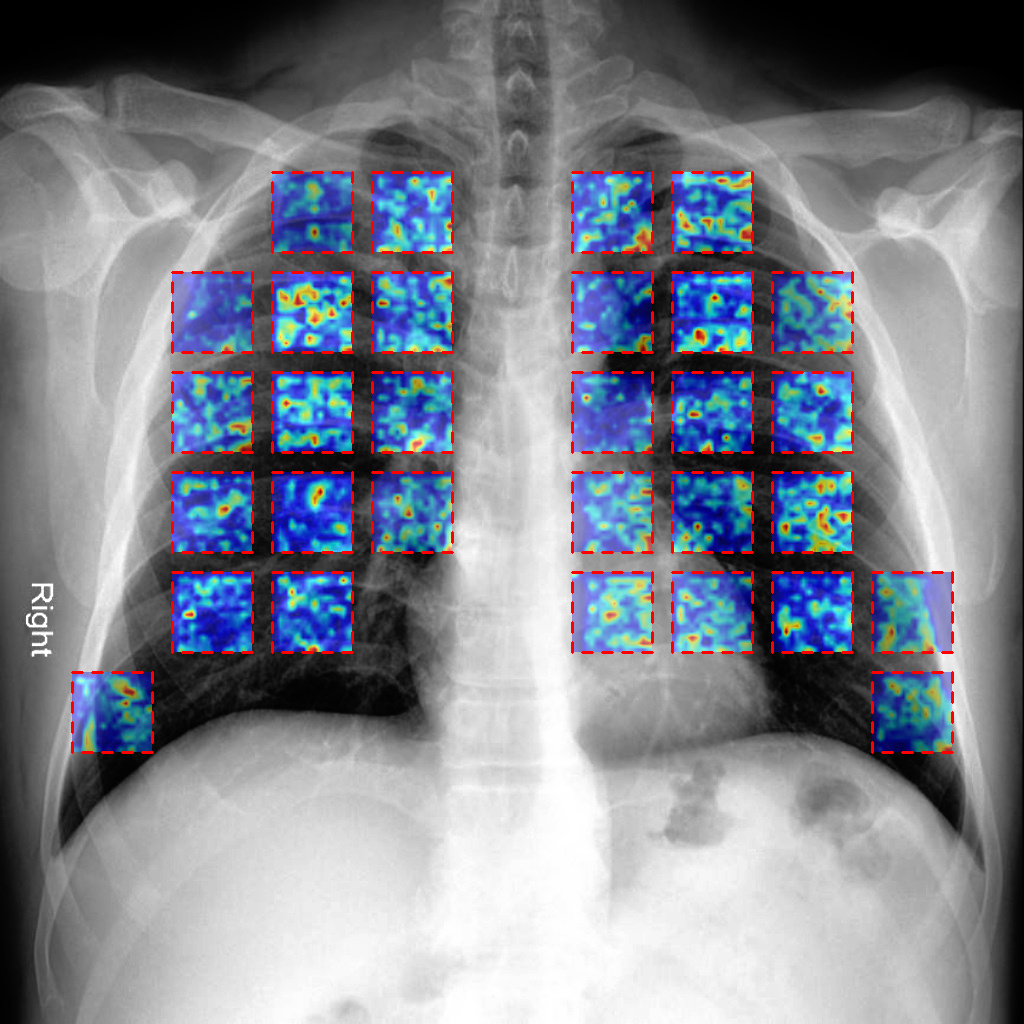} \\
        a) Wang \etal & b) Oh \etal & c) POTHER\\
    \end{tabular}
    \caption{The activation maps of: \textbf{(a)} cropped image global training, \textbf{(b)} segmented lung patch-based local training, \textbf{(c)} our multi-task patch-based local training method - \textbf{POTHER}. The globally trained model focuses on artificial electronics, textual information, and other non-disease-related features such as shoulder position. The model trained on the segmented lung is spared textual information but uses lung contour for classification. Our proposed model does not use these spurious features for classification but lung morphology instead.}
    \label{fig:globalgradcam}
\end{figure*}

\begin{figure*}[t!]
   \centering
\begin{tabular}{cc}
\includegraphics[width=4.7cm]{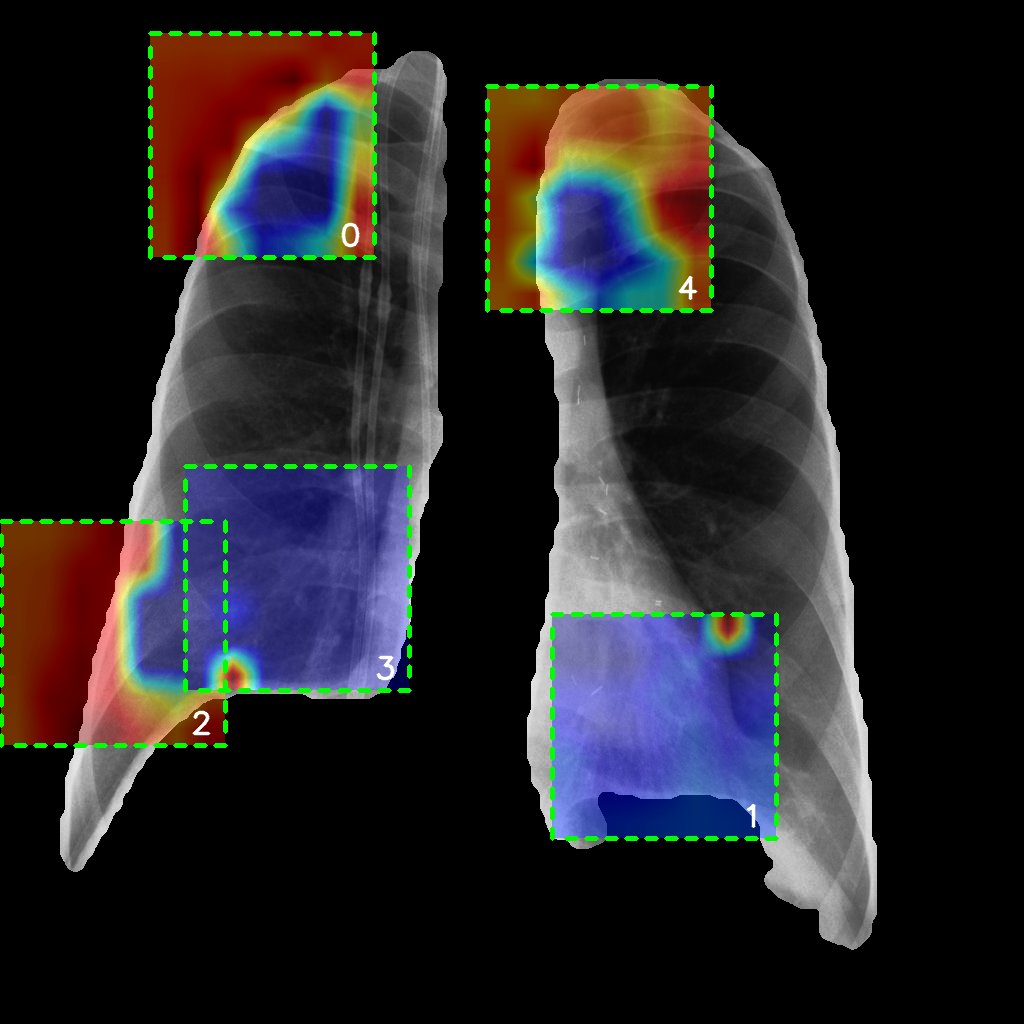}&
\includegraphics[width=4.7cm]{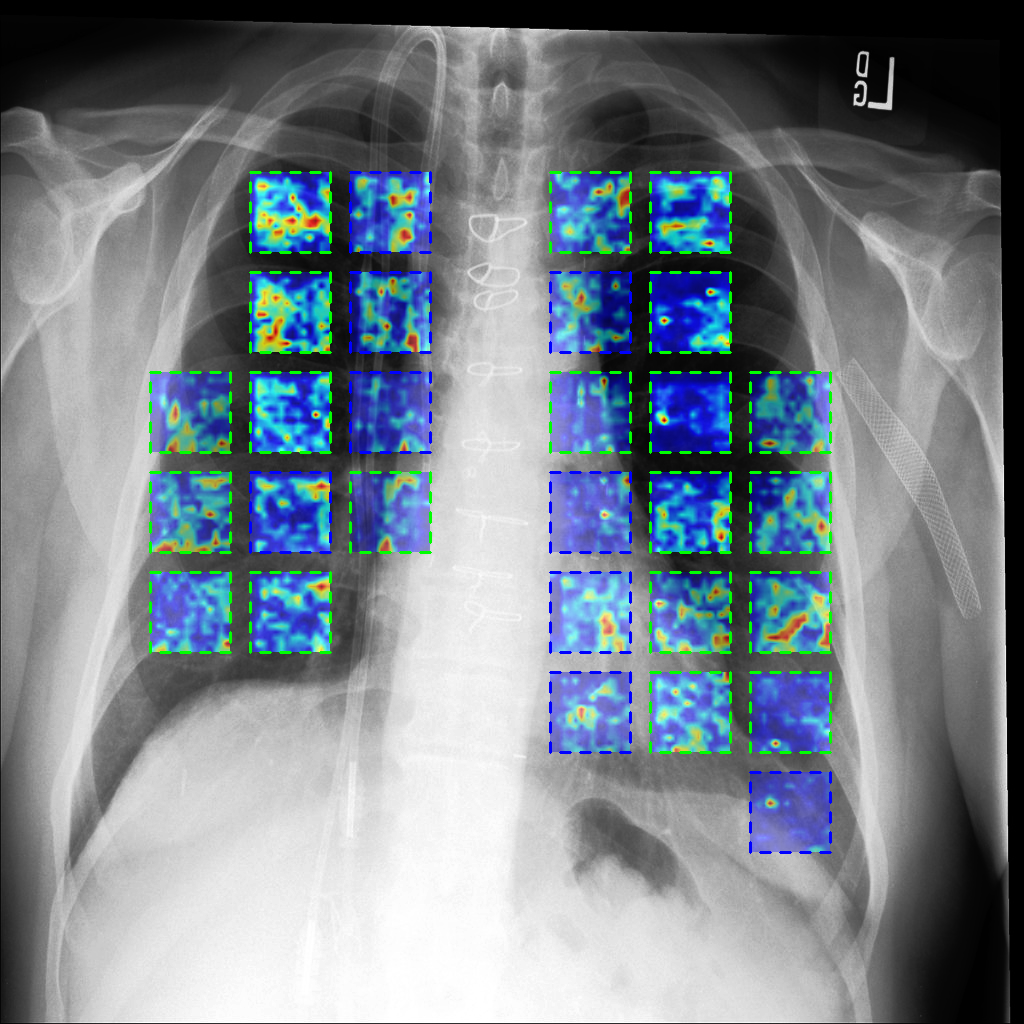} \\
\includegraphics[width=4.7cm]{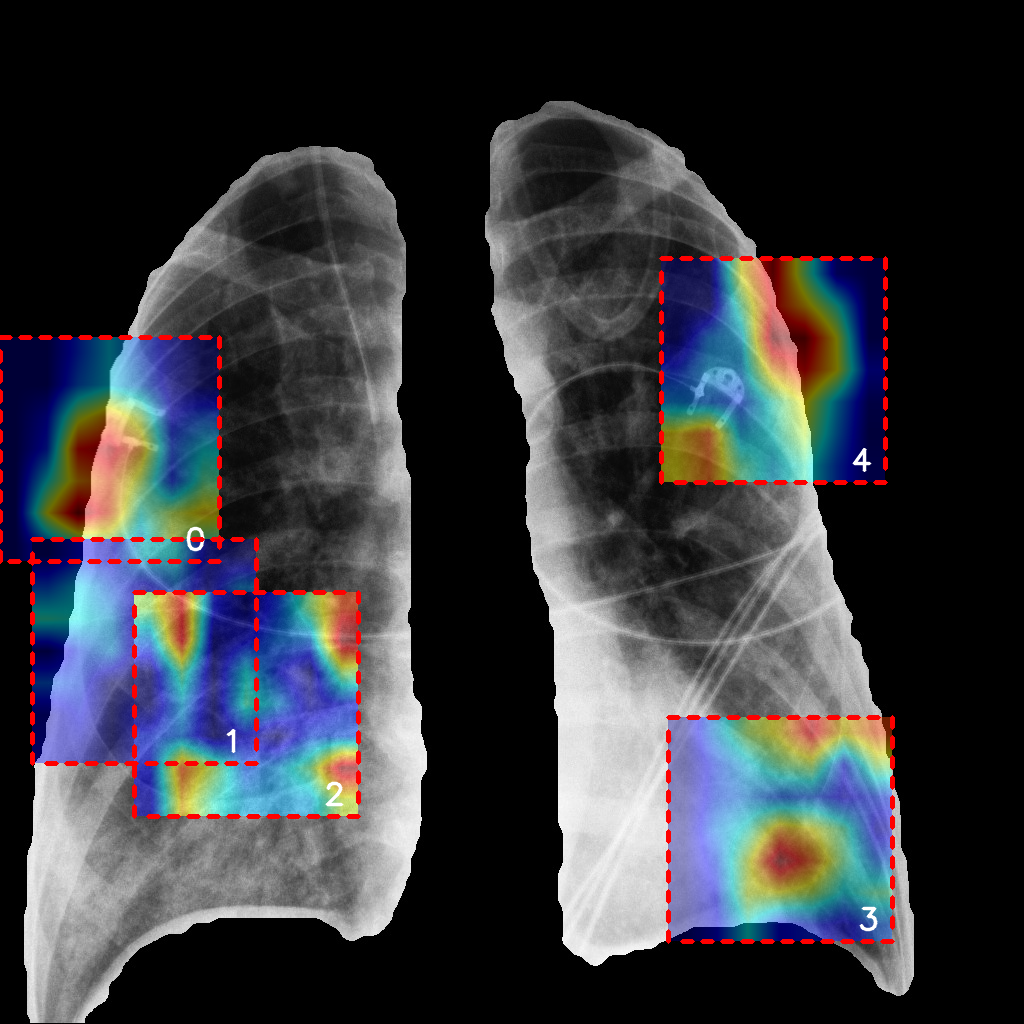}&
\includegraphics[width=4.7cm]{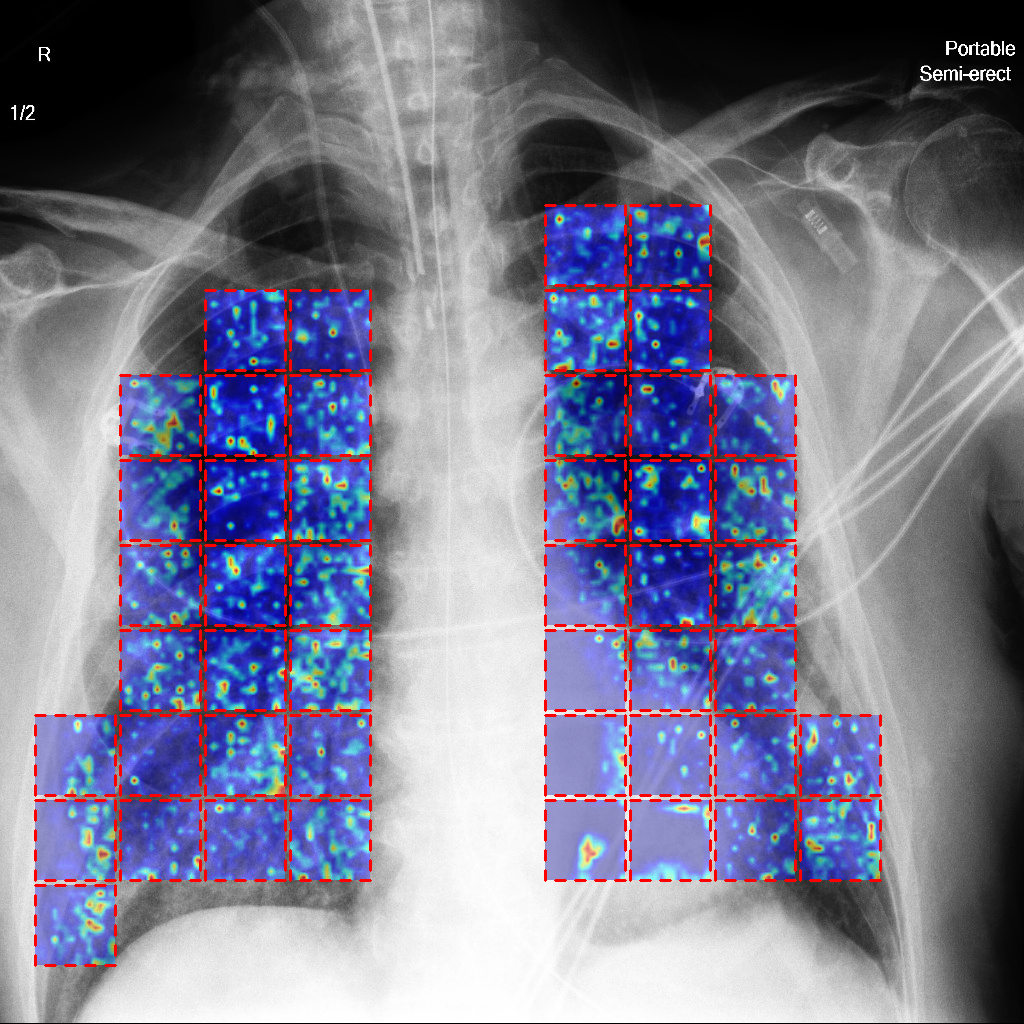} \\
a) Oh \etal & b) POTHER\\
\end{tabular}
    \caption{Patch learning activation maps comparison: \textbf{(a)} Oh \etal train the model using big patches which, if placed close to the lung edge, provide a model with awareness of lung contour, allowing it to learn that, (b) \textbf{POTHER} model focuses on fine-grained lung markings like, e.g. interstitial opacities.}
    \label{fig:ohvspother}
\end{figure*}

\begin{figure}[t!]
   \centering
\begin{tabular}{cc}
\includegraphics[width=4.5cm]{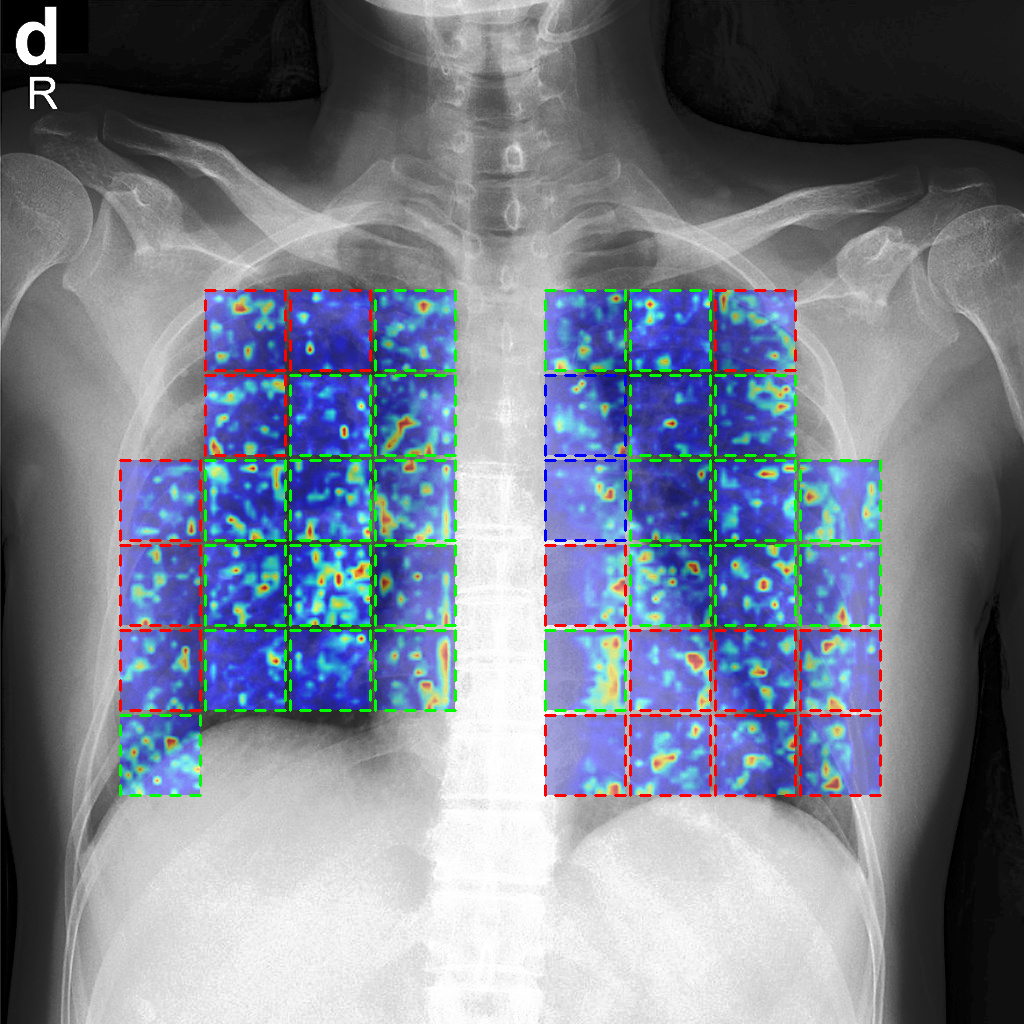}&
\includegraphics[width=4.5cm]{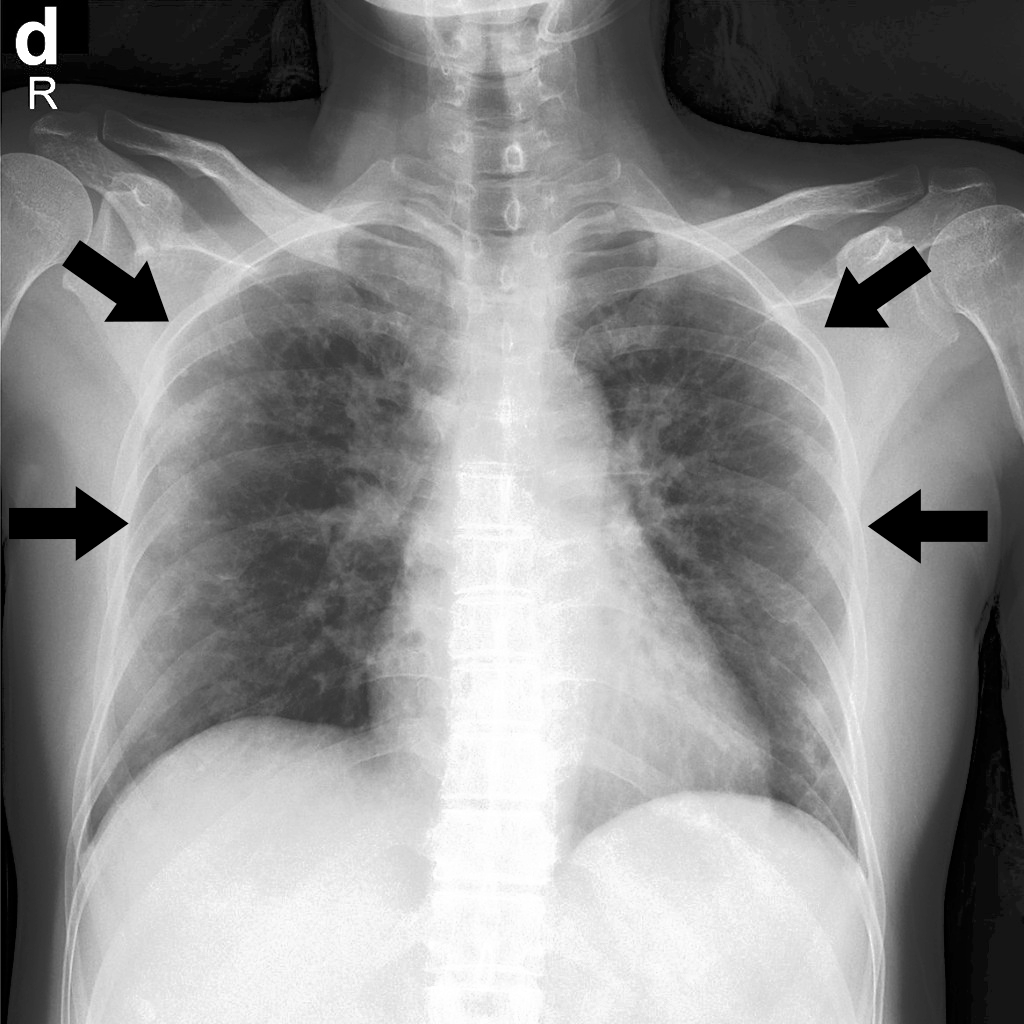} \\
\includegraphics[width=4.5cm]{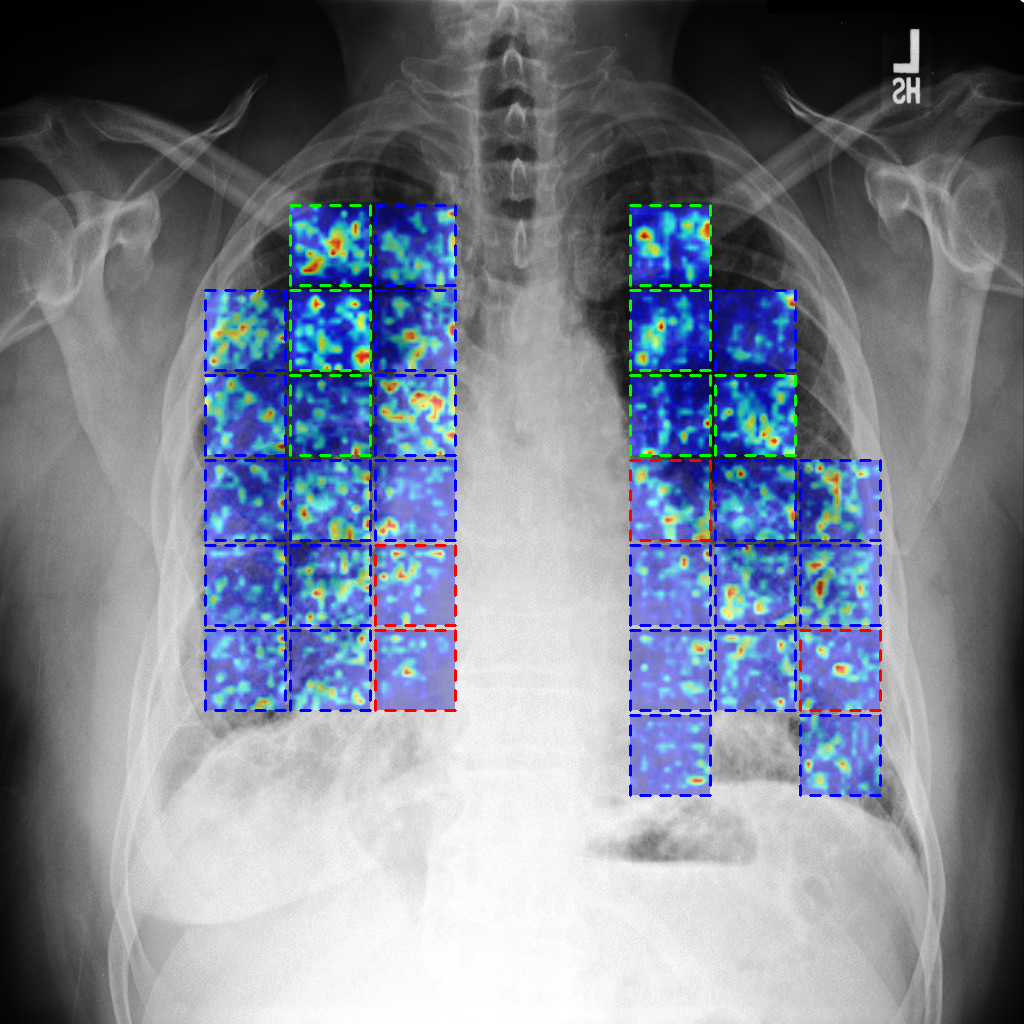}&
\includegraphics[width=4.5cm]{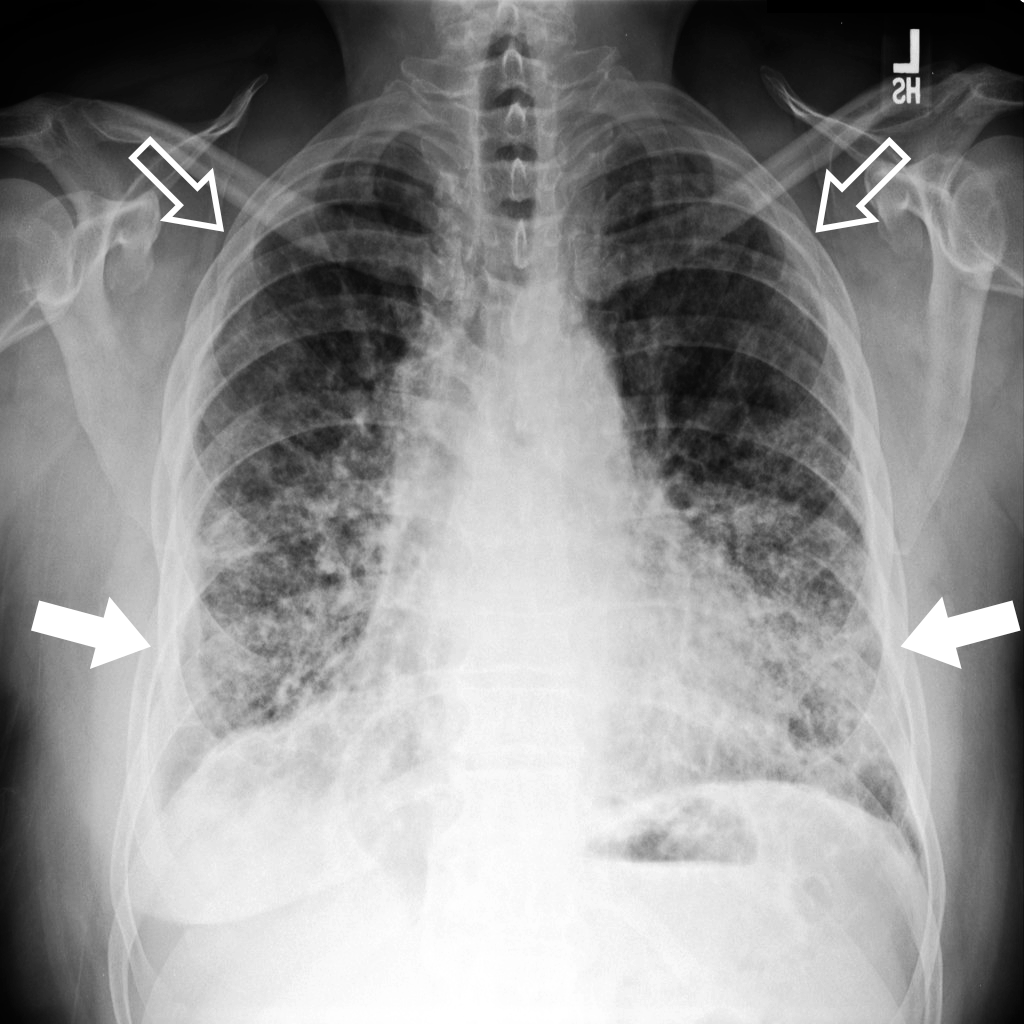} \\
a) Classification & b) Pathology area\\
\end{tabular}
    \caption{POTHER activation maps and per-patch classification results - \textbf{(a)} the model output class is coded with patch bounding box colour: \textit{green - normal lung, blue - pneumonia, red - COVID-19}. \textbf{(b)} black arrows point to the area with white opacity characteristic of COVID-19, white arrows point to lower lobes opacity characteristic to different types of pneumonia, outlined arrows point to the area free of opacity. The areas indicated by the arrows and the POTHER classifications overlap, indicating that the model may consider manifestations specific to the classified diseases when making decisions.}
    \label{fig:pothermesh}
\end{figure}

\begin{figure*}[t!]
    \centering
    %1
    \begin{subfigure}[t]{0.22\textwidth}
    \includegraphics[width=\textwidth]{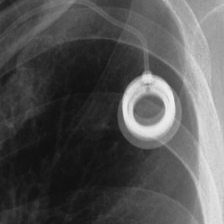}
    \subcaption{}
    \label{fig:fig5a}
    \end{subfigure}%
    \hspace{1em}%
    \begin{subfigure}[t]{0.22\textwidth}
    \includegraphics[width=\textwidth]{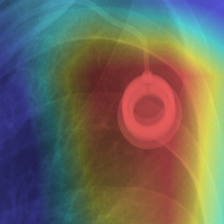}
    \subcaption{}
    \label{fig:fig5b}
    \end{subfigure}%
    \hspace{1em}%
    \begin{subfigure}[t]{0.22\textwidth}
    \includegraphics[width=\textwidth]{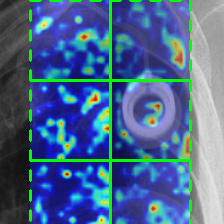}
    \subcaption{}
    \label{fig:fig5c}
    \end{subfigure}
    
    %2
    \begin{subfigure}[t]{0.22\textwidth}
    \includegraphics[width=\textwidth]{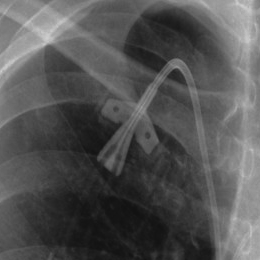}
    \subcaption{}
    \label{fig:fig5d}
    \end{subfigure}%
    \hspace{1em}%
    \begin{subfigure}[t]{0.22\textwidth}
    \includegraphics[width=\textwidth]{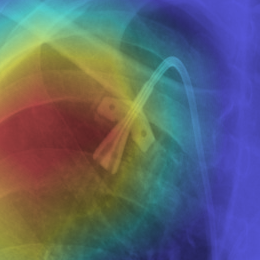}
    \subcaption{}
    \label{fig:fig5e}
    \end{subfigure}%
    \hspace{1em}%
    \begin{subfigure}[t]{0.22\textwidth}
    \includegraphics[width=\textwidth]{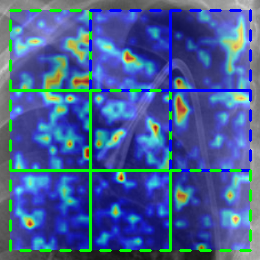}
    \subcaption{}
    \label{fig:fig5f}
    \end{subfigure}
    
    %3
    \begin{subfigure}[t]{0.22\textwidth}
    \includegraphics[width=\textwidth]{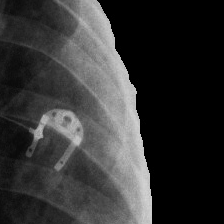}
    \subcaption{}
    \label{fig:fig5g}
    \end{subfigure}%
    \hspace{1em}%
    \begin{subfigure}[t]{0.22\textwidth}
    \includegraphics[width=\textwidth]{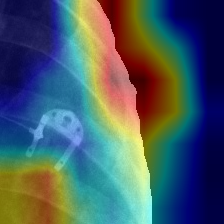}
    \subcaption{}
    \label{fig:fig5h}
    \end{subfigure}%
    \hspace{1em}%
    \begin{subfigure}[t]{0.22\textwidth}
    \includegraphics[width=\textwidth]{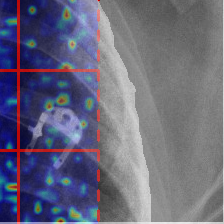}
    \subcaption{}
    \label{fig:fig5i}
    \end{subfigure}

    %4
    \begin{subfigure}[t]{0.22\textwidth}
    \includegraphics[width=\textwidth]{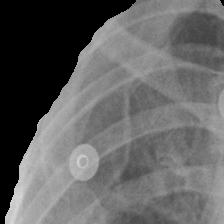}
    \subcaption{}
    \label{fig:fig5j}
    \end{subfigure}%
    \hspace{1em}%
    \begin{subfigure}[t]{0.22\textwidth}
    \includegraphics[width=\textwidth]{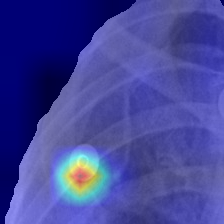}
    \subcaption{}
    \label{fig:fig5k}
    \end{subfigure}%
    \hspace{1em}%
    \begin{subfigure}[t]{0.22\textwidth}
    \includegraphics[width=\textwidth]{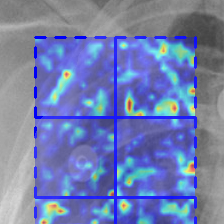}
    \subcaption{}
    \label{fig:fig5l}
    \end{subfigure}
    
   \caption{Analysis of activation maps in the vicinity of elements influencing the model decision. Comparison of the models: \textbf{(b, e)} trained globally, \textbf{(h, k)} segmented-lung patch-based and \textbf{(c, f, i, l)} with our \textbf{POTHER} method. The left column of figures \textbf{(a, d, g, j)} shows fragments of the original CXR image, figures \textbf{(b, e, h, k)} a fragment with superimposed activation map and the right column \textbf{(c, f, i, l)} activation maps of individual \textbf{POTHER} patches and their color-coded votes.}
   \label{fig:potherrobust}
\end{figure*}

We analyse probabilistic class activation maps for gradient weighted models that classify CXR images into three classes: normal, pneumonia and COVID-19 in the COVIDx dataset. It turns out that the globally \cite{COVID_Net} and the locally \cite{oh} trained model focus on non-disease related elements. In contrast, POTHER focuses on the morphological structure of lung tissue. The globally trained model concentrates on laterality tokens and ECG-leads artifacts even though input images are cropped. The locally trained model lacks textual information, but instead, it uses the lung mask's contour, supported by an edge's strong gradient because of the black background, to classify the image. The globally trained ResNet-50 model was fed with images pre-processed and augmented, as described in Wang's work. The pre-processing consists of cropping the top 8\% of the image, and according to the authors, it is to mitigate commonly-found embedded textual information, which, as presented in Fig. \ref{fig:globalgradcam}, is not enough because there are texts localised too close to the lung to be cropped.
\\ 
\indent
Fig. \ref{fig:ohvspother} compares activation maps of two models trained with a patch-based learning approach. We can see that Oh's patches cover a considerable percentage of the image, allowing the model to focus on the lung's corner or ECG leads. On the contrary, we train POTHER on the whole unsegmented images. It does not use pixels outside the lung because of small patches and reduced mask areas inside which patch centres can be located. Though some of the POTHER's patches are confused by cable on the healthy patient's image, thanks to their relatively small area, the vast majority of patches vote for a proper class during majority voting.
\\
\indent
We analyse activation maps of the POTHER model based on the visible manifestations of pneumonia and COVID-19 on CXRs. Visible with an untrained eye, pneumonia and COVID-19 pneumonia signs are related to increased lung density, which may be seen as whiteness in the lungs, which obscures the lung markings that are typically seen depending on the severity of pneumonia \cite{cleverley}. During COVID-19, markings are partially obscured by the increased whiteness. These can be subtle and require a confirmation from a radiologist. COVID-19 pneumonia changes, like lung involvement, are mostly bilateral on chest radiograph. Areas of lungs - Fig. \ref{fig:pothermesh}, where increased whiteness can be found are pointed with blue (pneumonia) and red (COVID-19) arrows, and the region free of it is marked with green arrows. It can be seen that POTHER's decisions about patches are correlated with them, which may suggest a classification based on the disease symptoms seen on the CXR.
\\
\indent
The locations and neighbourhoods of ECG lead elements and other non-anatomical artifacts are analysed in more detail. Fig. \ref{fig:potherrobust} presents generated activation maps of the analysed models, strongly indicating that electronic devices' presence in the image  focuses heavily the model's attention and can be an important factor in decision-making. Using POTHER, we classify the exact locations, and its activation maps show no significant focus on artificial elements. Fig. \ref{fig:fig5f} shows that 3 out of 9 POTHER's patches change their output near the cable; however, thanks to the majority voting it does not influence the final decision.

\subsection{Quantitative analysis}
To evaluate results, we use precision, recall, F1 score and accuracy. As a test set, we use a COVIDx introduced by Wang \etal \cite{COVID_Net}. It consists of 300 images equally distributed into three classes. A globally trained model’s inference results in a single classification per image, while inference of a locally trained one requires majority voting before making the final decision. When we train Oh's patch-based learning model, we follow the same pre-processing and augmentations as the author and the only differences are the model used for lungs segmentation and our mask filtration algorithm. We compare the performance of four models that are trained on the COVIDx dataset. The results presented in Table \ref{tab:quant_results} are comparable in terms of accuracy, which oscillates around 90\%. However, COVID-Net's accuracy, as well as its precision of COVID-19 classification, stands out. Our method achieves the highest F1 score of 0.974 for the COVID-19 class. Oh's model performs similarly to the results reported in their work, where it scored an accuracy of 0.889 on their dataset. 
\begin{table*}[ht!]\centering
    \caption{The performance comparison on COVIDx test set}
    \label{tab:quant_results}
    \ra{1.3}
    \begin{tabular}{@{}llccccccc@{}} 
        \toprule
        Method & Class & \phantom{aa} & Precision & Recall & F1 & \phantom{a} & Accuracy \\
        \cmidrule{1-2} \cmidrule{4-6} \cmidrule{8-8}
        ResNet-50 \cite{COVID_Net} & Normal    && 0.882 & 0.970 & 0.924 && \\
                                   & Pneumonia && 0.868 & 0.920 & 0.893 && 0.906\\
                                   & COVID-19  && 0.988 & 0.830 & 0.902 && \\
                                  
        COVID-Net \cite{COVID_Net} & Normal    && \textbf{0.905} & 0.950 & \textbf{0.927} && \\
                                   & Pneumonia && 0.913 & \textbf{0.940} & \textbf{0.926} && \textbf{0.933} \\
                                   & COVID-19  && 0.989 & 0.910 & 0.948 && \\
        Patch learning \cite{oh}   & Normal    && 0.815 & 0.970 & 0.886 && \\
                                   & Pneumonia && 0.914 & 0.813 & 0.860 && 0.886 \\
                                   & COVID-19  && 0.963 & 0.867 & 0.912 && \\
        \addlinespace 
        \midrule
        Ours (POTHER)              & Normal    && 0.790 & \textbf{0.980} & 0.875 && \\
                                   & Pneumonia && \textbf{0.963} & 0.780 & 0.862 && 0.903 \\
                                   & COVID-19  && \textbf{1.000} & \textbf{0.950} & \textbf{0.974} && \\
        \bottomrule
    \end{tabular}
\end{table*}
\section{Discussion} \label{discussion}
In this work, we propose a novel learning method specifically designed to address difficulties inherent in CXR images. In addition, we analyse the model's biases when it is making classification decisions. The improvement of diagnostic capabilities and the reduction of the negative influence of the numerous CBs in the COVIDx dataset guide us during model development. Many researchers have achieved very high scores on CXR images when trying to improve COVID-19 classification, but unfortunately, very few have analysed what allows the model to achieve such high scores.\\
\indent
When trying to improve COVID-19 classification based on CXRs, great attention should be paid to understanding what features allow the model to achieve good classification. Otherwise, we may obtain a model with little diagnostic value, though it achieves seemingly high scores utilising CBs to classify on dataset that it was trained and tested. We reveal potential pitfalls in the CXR medical imaging domain, i.e. some types of clues irrelevant to disease recognition, like hospital markings and ECG leads, causing the model to miss the point of the disease symptoms. The list of CBs is not exhaustive as it cannot  be guaranteed that the model will not shift its attention to something similarly undesired if known CBs are addressed.
Therefore, a dataset specifically designed to eliminate as many detected CBs as possible is desirable. Annotation or segmentation of the pathological lung area by a radiologist visible in the CXR would also be valuable. According to \cite{cleverley}, a chest radiograph can look normal in up to 63\% of people infected with COVID-19, especially in the early stages. Using COVID-19 labelled radiograph with no noticeable disease-related changes in the lung image is very likely confusing to the model and encourages learning of undesirable confounding features.\\
\indent

%We conclude that is a challenging dataset to create the model with diagnostic value based on COVIDx; and it is premature to report model trained on this dataset ready for medical use. Decisions about CXR images should be made only by trained radiologists or clinicians. The developed model can be treated only as an assistive tool to identify cases worthy of closer examination, but foremost as a further step in developing methods to create better models on CXRs, which will eventually prove a valuable aid to clinicians. Given the datasets already available, with their potential weaknesses in the form of CBs, we propose a method that reduces the impact of at least some of them, possibly improving the diagnostic value of future models trained with it.

\section{Conclusions} \label{conclusions}

% NOW:
% 1. what do we propose
% 2. what it does
% 3. why this task is difficult
% 4. why COVIDx is difficult
% 5. why pother copes with this difficulties
% 6. why multitask helps
% 7. why segmentation alone is not recommended
% 8. why pother is robust to CBs
% 9. plans 

%FIXED ARek notes
% In this paper, we proposed a novel multi-task patch-voted learning-based method called POTHER for CXR-image based COVID-19 detection. Our model learns pneumonia manifestations and lung morphology features. We perform the activation map analysis and show that the deep learning methods are susceptible to learning features unrelated to the pathology. We show that the COVIDx dataset contains many images that can confuse the learned model, such as ECG leads, laterality tokens or other hospital-specific markings. To this end, POTHER's patches include the area of lung fields and only its closes vicinity to eliminate the source of undesirable features. %
% %related to lung shape and its position in the global context.% 
% Using segmentation task with an attention mechanism provides an efficient feature extraction, allowing satisfactory results despite training with limited lung fragments. By contrast, we show that using segmented lung images as an input can be a source of inductive bias. Thanks to training with small patches, our method is less sensitive to the CBs. In future work, we will apply our method with Vision Transformer (ViT) \cite{ViT} on a large-scale CXR dataset to improve the model generalisation with more complex lung disease classes.

% ORIGINAL with Arek notes
In this paper, we proposed a novel multi-task patch-voted learning-based method called POTHER for CXR-image-based COVID-19 detection. Our model learns pneumonia manifestations and lung morphology features. We performed the activation map analysis and showed that the deep learning methods are susceptible to learning features unrelated to the pathology. The COVIDx dataset contains many images that can confuse the learned model, such as ECG leads, laterality tokens or other hospital-specific markings. To this end, POTHER's patches include the area of lung fields and only its closes vicinity to eliminate some of the sources of undesirable features. Using segmentation task with an attention mechanism provides an efficient feature extraction, allowing satisfactory results despite training with limited lung fragments. Thanks to training with small patches, our method is less sensitive to the CBs. In future work, we will apply our method with Vision Transformer (ViT) \cite{ViT} on a large-scale CXR dataset to improve the model generalisation with more complex lung disease classes.

\section*{Acknowledgements}

This work is supported in part by the European Union’s Horizon 2020 research and innovation programme under grant agreement Sano No. 857533 and the International Research Agendas programme of the Foundation for Polish Science, co-financed by the EU under the European Regional Development Fund. This research was funded by Foundation for Polish Science (grant no POIR.04.04.00-00-14DE/18-00 carried out within the Team-Net program co-financed by the European Union under the European Regional Development Fund), National Science Centre, Poland (grant no 2020/39/B/ST6/01511).

\end{document}